\documentclass[12pt, journal, onecolumn, compsocconf]{IEEEtran}

\usepackage{wrapfig}
\usepackage{geometry}
\usepackage{tipa}
\usepackage{times}
\usepackage{dsfont}
\usepackage{graphicx}
\usepackage[font=small,labelfont=bf,labelsep=period,tableposition=bottom]{caption}
\usepackage{amsmath}
\usepackage{amsfonts}
\usepackage{amsthm}
\usepackage{amssymb}
\usepackage{mathtools}
\usepackage{verbatim} % provides block comment
\usepackage{url}
\usepackage{hyperref}
\hypersetup{
    colorlinks=true,
    linkcolor=blue,
    filecolor=magenta,      
    urlcolor=cyan,
}
\usepackage[capbesidesep=quad]{floatrow}
\usepackage[tight]{subfigure}
\newfloatcommand{capbtabbox}{table}[][\linewidth]
\usepackage[export]{adjustbox}
\usepackage{courier}
\usepackage{slashbox}
\usepackage{hhline}
\usepackage{multirow}
\usepackage{booktabs}
\usepackage{colortbl}
\usepackage{enumitem}
\usepackage[dvipsnames]{xcolor}
\usepackage{balance}
\usepackage{array}

\definecolor{light-gray}{gray}{0.7}

\graphicspath{ {plots/} {figures/} {imgs/} }
\newcolumntype{R}[2]{%
    >{\adjustbox{angle=#1,lap=\width-(#2)}\bgroup}%
    l%
    <{\egroup}%
}
\begin{document}

% the solicitation says to put title in all caps
\title{Cloud Resource Optimization for Processing Multiple Streams of Visual Data}
\author{\IEEEauthorblockN{Zohar Kapach, Andrew Ulmer, Daniel Merrick, Arshad Alikhan, Yung-Hsiang Lu, \\}
\IEEEauthorblockA{
%School of 
% Electrical and Computer Engineering\\ 
Purdue University, West Lafayette, IN, USA\\\{{\tt zkapach, ulmera, dmerrick, aalikhan, yunglu\}@purdue.edu}} 

Anup Mohan% IEEEauthorrefmark{3} 

%\IEEEauthorrefmark{3}
Intel, Santa Clara, CA, USA \\
{\tt anup.mohan@intel.com}

Ahmed S. Kaseb% \IEEEauthorrefmark{4}
%\IEEEauthorrefmark{4}

Cairo University, Giza, Egypt\\
{\tt akaseb@eng.cu.edu.eg}

George K. Thiruvathukal% \IEEEauthorrefmark{1}\IEEEauthorrefmark{2} \\

% \IEEEauthorrefmark{1}
Loyola University Chicago, Chicago, IL, USA

% \IEEEauthorrefmark{2}
Argonne National Laboratory, Argonne, IL, USA\\
{\tt gkt@cs.luc.edu} \\
% \IEEEauthorblockA{\IEEEauthorrefmark{1}Department of Computer Science\\ Loyola University Chicago, Chicago, IL, USA\\
% {\tt gkt@cs.luc.edu}}
}

\maketitle 
\thispagestyle{empty}

%\tableofcontents

% TODO: avoid the "forward ref" in model char.
% TODO: figure 2a to figure 1 (understand crop)
% TODO: figure 1 -> figure 2

% \input{cover_letter}
% \clearpage
\begin{abstract}
Hundreds of millions of network cameras have been installed throughout the world. Each is capable of providing a vast amount of real-time data. Analyzing the massive data generated by these cameras requires significant computational resources and
the demands may vary over time. Cloud computing shows the most promise to provide the needed resources on demand. In this article, we investigate how to allocate cloud resources when analyzing real-time data streams from network cameras. A resource manager considers many factors that affect its decisions, including the types of analysis, the number of data streams, and the locations of the cameras. The manager then selects the most cost-efficient types of cloud instances (e.g. CPU vs. GPGPU) to meet the computational demands for analyzing streams. We evaluate the effectiveness of our approach using Amazon Web Services. Experiments demonstrate more than 50\% cost reduction for real workloads.

\begin{comment}
Companies such as Microsoft, Amazon, and IBM offer a variety of cloud-based resources to address this and similar problems, but no method to tailor these resources for a specific user{'}s needs. Many optimizers have been developed to determine the best configuration of these cloud-instances for a singular need, but there is no solution that reduces cost and boosts productivity across the board. We propose a resource manager that considers both location and type when determining the optimal combination of cloud-instances. Consequently, our method has been proven to maximize efficiency and minimize cost when using cloud-instances for large-scale, real-time, computer vision.We evaluated our method with video data that varied in frame rate, frame size, and location. Using this process to assess the resource manager, we establish consistency across the range of possible inputs. This resource manager has been shown to save up to 60\% more on cloud-instance cost than other popular optimizers, without sacrificing performance.	
\end{comment}

\end{abstract}

%\begin{IEEEkeywords}

%\end{IEEEkeywords}

\vspace{-.3cm}

\section*{Introduction}\label{sec:introduction}

Visual data is projected to grow exponentially in the coming years. Video is expected to grow 26\% annually and account for 82\% of consumer Internet traffic; live video on the Internet is expected to grow 1,500\% from 2016 to 2021~\cite{CiscoVisualNetwork}. Surveillance cameras are expected to see similar growth. Today, more than 240 million surveillance cameras have been installed globally~\cite{JenkinsIHS2015} and Stratistics MRC predicts that the market will continue to grow 18.3\% annually. These video surveillance cameras can produce vast amounts of real-time data. With the rapid progress of machine learning and computer vision, it is possible to analyze these real-time streams. Two key technologies are essential to harness the potential of these real-time data steams: (1) analyzing the data using computer vision, and (2) employing scalable resources to meet the demands of computation.

This paper focuses on solving the second problem. This research group adopts both supercomputing and cloud computing to address the problem from different perspectives. Supercomputing systems are optimized for speed and I/O throughput but are limited when it comes to meeting fluctuating demands because of job scheduling requirements. Cloud computing offers the potential for high-end computing resources and also allows for on-demand operation. Proponents of cloud computing are willing to make trade-offs when it comes to exchanging CPU and I/O speed in favor of high availability and flexible scheduling. This paper describes our experiences with cloud computing to analyze many video streams from network cameras.

Many factors affect the resource requirements for analyzing visual data streams, including the complexity of the analysis programs, the content being analyzed, the size (number of pixels) of each image or video frame, the frame rates, etc.  The requirements may vary over time. For example, a program that analyzes video streams from traffic cameras to detect congestion may  run during rush hours only. Cloud computing is the best option to meet these varying needs.
Cloud computing allows users to dynamically allocate virtual machines (called {\it instances}) on demand.  Cloud vendors, such as Amazon EC2 (Elastic Cloud Computing), Microsoft Azure, and IBM Cloud, provide many types of cloud instances with different amounts of memory, number of cores, and number of GPUs (graphics processing units) at different prices (US dollars per hour). These instances reside in data centers distributed in North and South America, Europe, Asia, and Australia. The challenge is minimizing the cost of cloud instances without sacrificing performance.
   
To drive the research in cloud resource optimization for processing multiple visual data streams, a software infrastructure named {\it Continuous Analysis of Many Cameras} (CAM$^2$) has been established at Purdue University~\cite{7416927}. CAM$^2$ uses network cameras that provide real-time visual data publicly available on the Internet. These cameras observe traffic intersections, metropolitan areas, university campuses, tourist attractions, etc. Some cameras provide videos and others show snapshots. The analysis of real-time data streams can be used in many applications, such as urban planning and emergency response.\newline

\begin{comment}

aims to address this dilemma by providing a platform capable of not only pooling thousands of network cameras into a ready-to-use database, but also by leveraging recent developments in cloud-computing to effectively deploy computer vision methods on real-time video data. There is undoubtedly a colossal number of applications for this infrastructure, the most striking of which may be emergency response. As the world population grows and weather patterns become more extreme, society demands a robust response capacity now more than ever. But CAM$^2$ is more than simply the first steps towards this and similar capabilities, it lays the foundations for any system seeking to make sense of the now massive amount of information we have at our fingertips.

\end{comment}

\begin{tabular}{|p{6in}|} \hline
\\ 
%\multicolumn{1}{|c|} {\bf CAM$^2$ Sidebar} \\

CAM$^2$ is designed as a computing platform for analyzing real-time network camera data. Network cameras capture all sorts of visual data. The examples below show a night club, a computer classroom, a park, and a city street. 
CAM$^2$ has the following major components: (1) a camera database storing the information about the cameras' geographical locations, frame rates, etc, (2) a run-time system retrieving data from the network cameras and sending the data for analysis, and (3) a resource manager allocating and releasing computing resources to meet application demands.  CAM$^2$'s application programming interface (API) allows the same analysis programs to retrieve and analyze data from a diverse set of network cameras~\cite{7032135}. CAM$^2$ uses only public data available on the Internet. 
Readers interested in using CAM$^2$ can register at {\tt https://www.cam2project.net}. 
The Terms of Use and Privacy Policy are available on the website.
\\ \\

 {\includegraphics[height=0.87in]{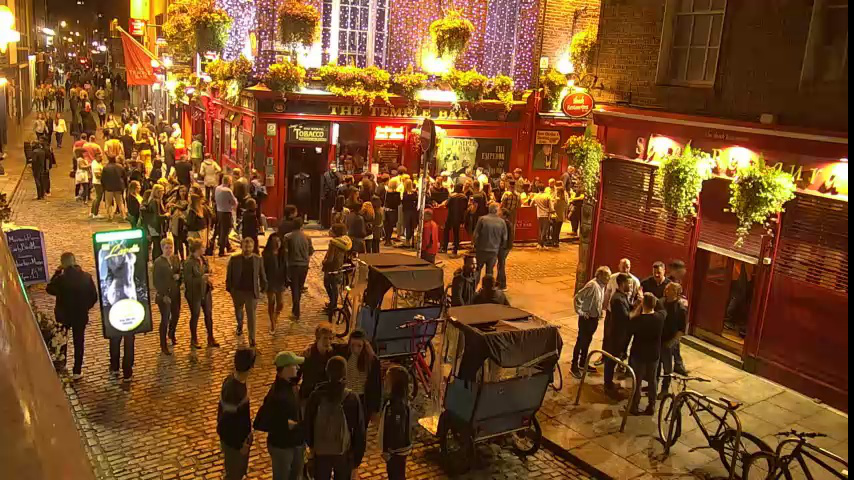}}
  {\includegraphics[height=0.87in]{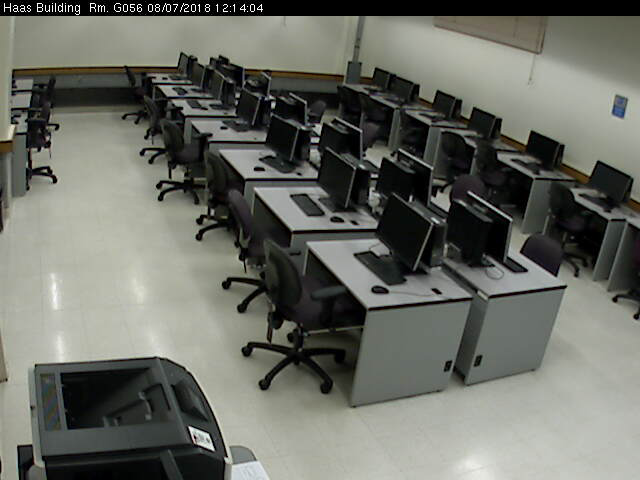}}
  {\includegraphics[height=0.87in]{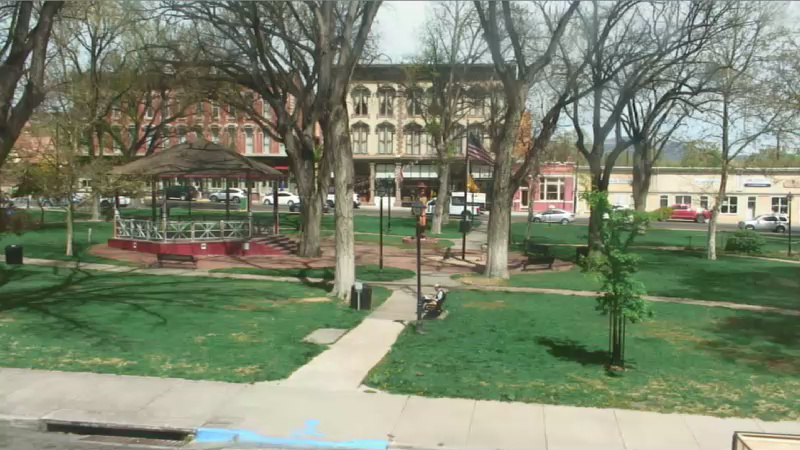}}
   {\includegraphics[height=0.87in]{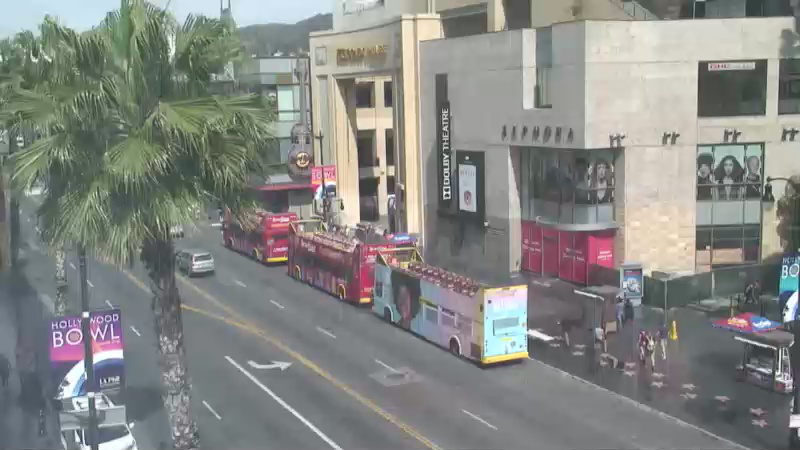}}
 
\\ \hline
\end{tabular}

\section*{Resource Management of Cloud Instances}
\label{sec:resource_man} 

As mentioned earlier, cloud instances would be ideal solutions for meeting the varying demands of video analytics. Cloud vendors provide many options; some instances have GPUs while the others do not. Amazon EC2 has optimizers for processes, memory, and networks. IBM cloud gives users the option of selecting virtual machines or physical machines (called ``bare metal servers''); within each category there are dozens of options with different types of processors and amounts of memory. Microsoft Azure also has many configurations to choose from.  Table~\ref{table:cloudprice} shows the prices of several types of cloud instances at different locations.   

 \begin{table}[h]
\begin{tabular}{|l|l|r|r|r|r|r|r|} \hline
Vendor                   & Instance                 & Cores & Memory (GiB) & GPU  & \multicolumn{3}{c|}{Price Per Hour (US\$)}                                                        \\ \hline
\multicolumn{5}{|l|}{}                                                                                                  & Virginia & London & Singapore \\ \hline
\multirow{3}{*}{EC2}   & c4.2xlarge               & 8                         & 15                               & 0                        & 0.398                        & 0.476                      & 0.462                         \\
                         & c4.8xlarge               & 36                        & 60                               & 0                        & 1.591                        & 1.902                      & 1.848                         \\
   & g3.8xlarge               & 32                        & 244                              & 2                        & 2.280                        & N/A                        & 3.340                         \\ \hline
\multicolumn{5}{|l|}{}   & US East                      & West Europe                & East Asia    \\ \hline
\multirow{2}{*}{Azure} & D8 v3                    & 8                         & 32                               & 0                        & 0.384                        & 0.480                       & 0.625                         \\
 & NC24r                    & 24    & 224          & 4    & 3.960    & 5.132  & N/A      
\\ \hline
\end{tabular}
\caption{Prices of different Amazon EC2 and Microsoft Azure cloud instances at different locations.}
\label{table:cloudprice}
\end{table}

\begin{figure}[ht]
    \centering
    \includegraphics[width=5in]{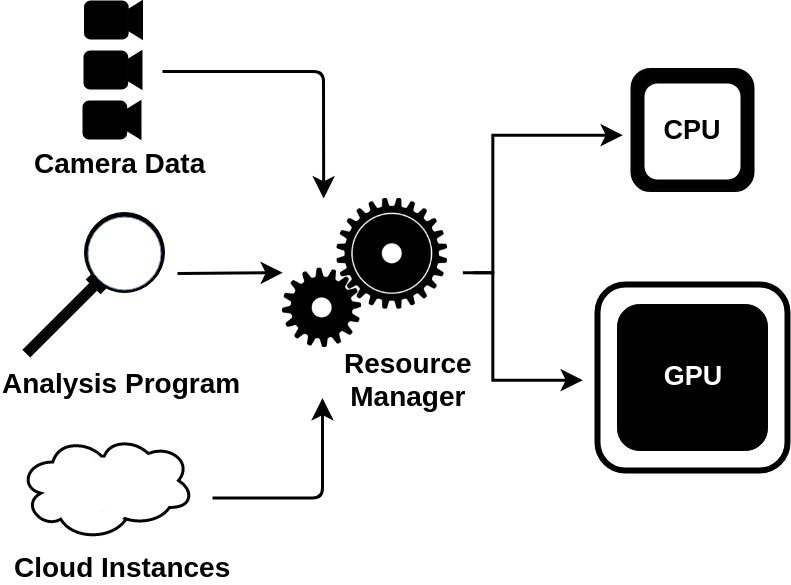}%
    \caption{The resource manager considers many factors, including the data content, the analysis programs,
 and the types and costs of cloud instances. The manager selects the most cost-efficient instances (these may have CPUs only or they may include GPUs as well).}
    \label{fig:cam2_system}
\end{figure}
Because of the wide range of prices, instance types, and locations, a resource manager is essential when optimizing large scale cloud usage. Figure~\ref{fig:cam2_system} illustrates the role of a resource manager. The resource manager has to consider the following factors when selecting  the most appropriate cloud instances: 

\begin{itemize}
\item \textbf{Characteristics of analysis programs.} Different programs have different resource requirements: some programs benefit from more cores, some need more memory, and some need GPUs.

\item \textbf{Desired frame rates.} Some analysis programs (such as tracking moving objects) need high frame rates. For some other programs (such as observing air quality or traffic congestion), low frame rates are sufficient.  

\item \textbf{Image or frame sizes, in terms of pixels.} If an image has more pixels, more computation is needed.

\item \textbf{Content of the data.} The execution time and resource requirements depend on the complexity of the content. Complex content (e.g., many moving objects in a video stream) may require more computing resources than simple content. 

\item \textbf{Locations of cameras and cloud instances.} When analyzing video streams, the distances between cloud instances and cameras (measured by the round-trip time) can affect the frame rates~\cite{7331200}. Therefore, there may be restrictions on the location of an instance.
\end{itemize}

Based on these inputs, the resource manager selects the cloud instances. An instance's configuration (also called the type) corresponds to the number of cores, the amount of memory, the presence of GPUs, and the geographical location. The choice is made to meet the resource requirements at the lowest possible cost. This resource manager is dynamic and its decisions may change over time because the demands may vary. 	This paper presents two optimization strategies: one manages CPU and GPU usage, and the other manages the locations of instances.
    
\vspace{0.1in}    
\begin{tabular}{|p{6in}|} \hline
\\ \multicolumn{1}{|c|}{\bf Cloud Resource Optimization Problem} \\

The problem: selecting the cloud instances (types and locations) for meeting the resource requirements needed to analyze real-time streaming data at the lowest costs.

\\ \hline
\end{tabular}
\vspace{0.1in}

\section*{Adaptive Resource Management for Video Analysis in the Cloud}
A solution to handling cloud resource management, proposed by Mohan et al. \cite{mohan2018adaptive}, is Adaptive Resource Management for Video Analysis in the Cloud (ARMVAC) . This method does the following: (1) reads inputs necessary for modeling the problem as a Vector Bin Packing Problem, (2) selects the locations of cloud instances to be considered for the given analysis, (3) determines the types and number of cloud instances needed for the analysis, and (4) employs an adaptive resource management solution to adjust resource requirements during runtime.  Kaseb et al. \cite{kaseb2018analyzing}  improve
ARMVAC by considering cloud instances with both CPU and GPU.  Mohan et al. \cite{mohan2016location} extend ARMVAC by considering instances' locations.
The following sections explain these improvements.
% Although a relatively thorough approach, ARMVAC has a variety of shortcomings. Most notably, when framing the problem as a Vector Bin Packing Problem, ARMVAC fails to account for both CPU and GPU usage, as well as the cost of using instances that have both types of processing units - a problem addressed byand elaborated on below. ARMVAC also neglects to balance the cost of a cloud-instance with its location - another issue addressed byand elaborated below.

\section*{CPU and GPU Management in the Cloud}
\label{sec:cpu_gpu_optim} 

	\subsubsection*{\textbf{Differences between CPUs and GPUs}}Central processing units (CPUs)  and graphics processing units (GPUs, also called general-purpose 
    graphics processing units GPGPUs) have different characteristics and capabilities. CPUs, with several to dozens of processing cores, are the brains of computers. CPU cores can handle complex program flows with many control statements. In contrast, GPUs have thousands of smaller and simpler cores. GPUs can be significantly faster when performing similar tasks on many pieces of data, such as videos and images. GPUs adopt the SIMD (single instruction, multiple data) style parallelism: the same computation runs on different elements of arrays. 

	Another key difference between CPUs and GPUs is price. For example, Amazon EC2's {\tt c5d.9xlarge} CPU instance has 36 virtual CPUs with 72 GB of memory and costs \$1.728 per hour. GPUs, however, tend to be much more expensive. The {\tt p3.2xlarge} GPU instance has 8 virtual CPUs with 61 GB of memory and costs \$3.06 per hour. Another GPU instance, {\tt p3.8xlarge}, has 32 virtual CPUs and 244 GB memory and costs \$12.24 per hour.

\begin{tabular}{|p{6in}|} \hline
\label{packing_problem_sidebar}
\\ \multicolumn{1}{|c|}{\bf Multi-Dimensional Multi-Choice Packing Problem} \\
\parbox{1 cm}\indent 
Multi-dimensional multi-choice packing problems occur in everyday life. Consider renting a truck (or trucks) for moving boxes. The trucks come in different sizes at different costs and the boxes have different dimensions (length, width, height) as well. The objective is to find the cheapest truck (or trucks) that can accommodate these boxes. Some boxes, however, have more requirements than others. For example, it is possible that a box may include frozen food and would need a truck that offers refrigeration.\newline
\parbox{1 cm}\indent This problem is analogous to the problem of finding the most cost-effective cloud instances for analyzing the data streams. Each type of instance can be compared to a type of truck, in that it has several dimensions: the number of CPUs, the amount of memory, and the presence of GPUs. Each analysis program, running on one data stream, is a box with a particular size. Some analysis programs (e.g., tracking) need GPUs to process the data at the desired frame rates, similar to the boxes that require refrigerated trucks. Some other programs can use low frame rates and can run on CPUs only. Finding the most cost-effective cloud instance (or instances) to accommodate the analysis programs for all data streams is comparable to finding the most cost-effective truck (or trucks) to transport all boxes.\\

\parbox{1 cm}\indent In order to explain the multiple choice vector bin packing problem, we created an arc-flow graph like the one presented by Brandao and Pedroso \cite{brandao2016bin}. In the arc-flow graph illustrated below, the nodes weight represents the box dimensions (width, height), and the demand represents the number of boxes. Any path from the source node to the target node represents a viable set of boxes to pack into a truck. In this example, a truck with dimensions (7, 3) is filled with 3 types of boxes with the following weights and demands:
\\
\begin{enumerate}
    \item[A] Weight: (5, 1)  Demand: 1
    \item[B] Weight: (3, 1)  Demand: 1
    \item[C] Weight: (2, 1)  Demand: 2
\end{enumerate}
\\
First, box A is added as many times as the demand requires without over filling the truck. Then, box B goes through this process. And finally, box C. Once this graph is built, a second step is required to compress the graph. In cases where there can be hundreds of boxes and hundreds of trucks, a compressed graph is required to reduce the number of paths using the same set of boxes. This in turn will result in time saved when solving the graph. After the compression is complete a Gurobi 5.0.0 branch-and-cut solver is used to find the best path to get the maximum number of boxes into a truck. This method, however, only solves for one type of truck. In order to solve for multiple choices of trucks, we implemented the multiple choice method used by Brandao and Pedroso \cite{brandao2013multiple}. In this implementation, a graph is constructed for each truck type, and then solved using the Gurobi solver.
\\
    \begin{center}
     {\includegraphics[width=5in]{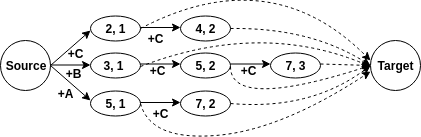}}
    \end{center}
    
\\ \hline
\end{tabular}

\vspace{0.1in}

\clearpage

\begin{figure}[h]
	\centering
    \subfigure[]
    {\includegraphics[width=3in]{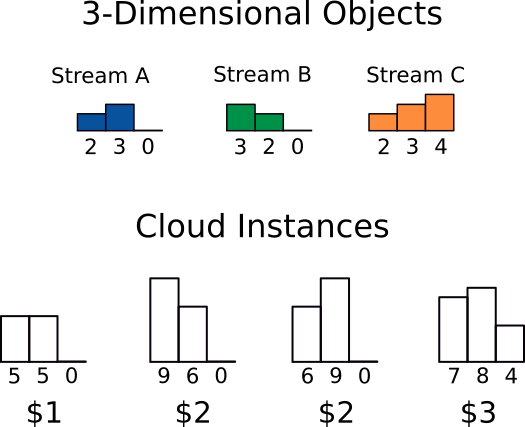}}%
    
    \subfigure[]
    {\includegraphics[width=4in]{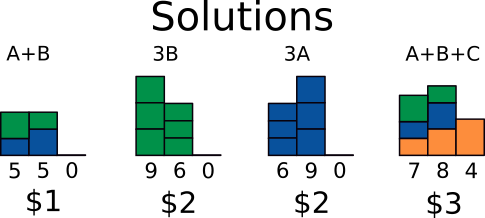}}%
    \caption{(a) A three-dimensional multiple-choice packing problem. A - C represent three types of data streams whose resource requirements are expressed in three dimensions (CPU, Memory, and GPU). The heights represent the resource requirements, comparable to the width, height, and lengths of boxes. There are four choices of cloud instances each offering different sizes and costs. They are comparable to different types of trucks. The heights represent the sizes of these dimensions, comparable to the sizes of trucks. The goal is to fit the data streams (boxes) into the cloud instances (trucks).
    (b) Four possible solutions to accommodate different combinations of data streams. 
    }
    \label{fig:bin_packing_problem}
    \label{fig:bin_packing_solution}
\end{figure}

	\subsubsection*{\textbf{Select Instances with CPU and GPU}}The significant differences in price and performance of CPU-only and GPU-equipped instances makes it critical to select the most cost-effective instances when analyzing many real-time data streams. To effectively select CPU and GPU instances for this task, Kaseb et al.~\cite{kaseb2018analyzing} formulated the problem as multi-dimensional multi-choice packing problem (please see the sidebar for explanation). This is illustrated in Figure~\ref{fig:bin_packing_problem} (a). In this figure, there are three types of data streams and four types of cloud instances. The goal is to fit the streams so that they completely fit inside the cloud instance and as little space is wasted as possible. Figure~\ref{fig:bin_packing_solution} (b) shows possible solutions to effectively pack different combinations of the data streams.

% Though the cloud instances in Figure~\ref{fig:bin_packing_solution} seem optimized, they are in fact not. When testing solution, Kaseb discovered that over-utilizing the CPUs and GPUs reduced the analyzing applications performance. Hence, in order to properly optimize and instance, only 90\% of an instance should be used.

Kaseb's solution organizes the resource requirements into four dimensions: CPU, memory size, GPU, and GPU memory size. The method considers the frame sizes and frame rates of the video streams for determining the resource requirements needed to run analysis on different data streams. Due to the fluctuations in executing the analysis programs, the study discovers that when any dimension is more than 90\% utilized, the performance starts to degrade. Thus, the method keeps the utilization of each dimension below 90\%. This multi-dimension, multi-choice optimization solution demonstrates considerable cost savings in different experimental settings. 

\begin{figure}[h]
  \centering
  {\includegraphics[width=6.5in]{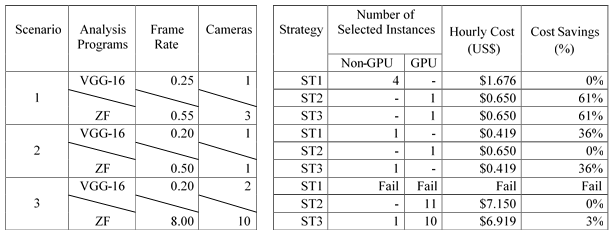}}
  \caption{Consider three scenarios and three unique instance selection strategies. Each scenario runs two analysis programs, VGG16 and ZF, at a different combination of frame rates and number of cameras. Strategy 1 (ST1) uses instances with only CPUs, strategy 2 (ST2) uses instances with only GPUs, and strategy 3 (ST3), Kasab's method, selects between GPU and CPU instances.  }
  %I believe the last sentence is necessary since it explains the strategies, without this sentence, I feel the strategy column seems confusing.
  \label{fig:costSavings_CPUGPU}
\end{figure}

Evaluation results from \cite{kaseb2018analyzing} are shown in Figure \ref{fig:costSavings_CPUGPU}. The experiments use ten network cameras from CAM$^2$'s database, with frame rates varying from 0.2 frames per second to 8 frames per second. Two  object detection programs are used to analyze the data: VGG16 \cite{simonyan2014very} and ZF \cite{zeiler2014visualizing}. At the highest frame rates, GPUs can accelerate these two analysis programs up to 16 times. At the lowest frame rates, the improvement falls below 5\%. In other words, the benefits of GPUs are apparent only when the frame rates are high. At low frames rates, CPUs are preferred because of the lower costs. This solution can reduce the costs by as much as 61\% by matching the resource requirements of the analysis programs and the cloud instances' capabilities. 

This study provides deep insights on how to offer computer vision services at lower costs as they become widely available in the cloud.  This study, however, does not consider the geographical locations of network cameras. The next section explains how cameras' locations impact network distances and the cloud resource management of different types of instances.
\label{results_CPU}

\section*{Optimizing Instance Type and Location}
\label{sec:loc_based_optim} 

\label{location_section}
In order to determine the most effective configuration of resources, the resource manager considers the cost of an instance in the context of its location. To make these considerations in practice, the location and type are first evaluated independently as follows.

% {\bf Zohar: this section's structure is different from the previous. The previous section does not have
% subsubsection. This section does. Why?
% }

\subsubsection*{\textbf{Local Optimization}}
Table~\ref{table:cloudprice} shows that the same type of cloud instance at different locations can have different costs. Sometimes this cost disparity can exceed 60\%. For example, the Azure {\tt D8 v3} instance costs 63\% more in Singapore than in Virginia USA ($\frac{0.625}{0.384} = 1.63$). A natural question is whether the data from network cameras should be sent to the cloud instances with the lowest prices. A prior study~\cite{7331200} shows that the observed frame rate is reduced when the distance (measured by 
network round-trip time) between a network camera and a cloud instance increases.  Thus, when analyzing data streams from worldwide network cameras, the locations of the cameras and cloud instances must be considered.

\begin{figure}[h]
  \centering
  \subfigure[]
  {\includegraphics[width=5in]{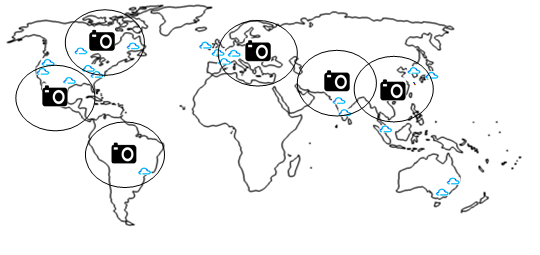}}
  \subfigure[]
  {\includegraphics[width=5in]{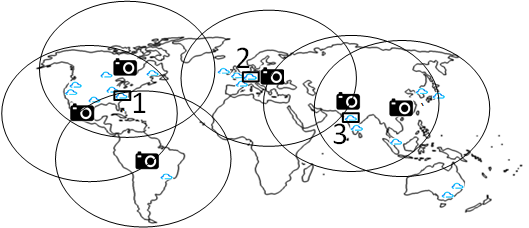}}
  \caption{Consider six network cameras geographically distributed in America, Europe, and Asia. Amazon EC2 offers cloud instances in many locations (shows as the cloud icons). In (a), if high frames are required, the data must be analyzed by instances that are closer to the cameras. The circles mark the maximum distances the data can travel. As a result, six instances are needed. In (b), the required frame rates are lower and the circles are larger. One cloud instance can analyze multiple data streams and only three instances are needed. The three selected instances, marked with black boxes, are one potential solution.  }
  \label{fig:cameras_over_map}
\end{figure}

Existing video cameras are designed for human viewing. For this purpose, 30 frames per second provide seamless experience. When video streams are analyzed by computers, the needed frame rates depend on the purposes. Though high frame rates are needed for tracking fast moving objects, low frame rates are sufficient for observing phenomena such as weather. Mohan et al.~\cite{8397037} study the necessary frame rates to track objects such as people walking, jogging, cycling etc. The study discovers that for cameras watching pedestrians walking, the frame rates can be reduced to as low as six frames per second. For objects that are far away from the cameras, even lower frame rates suffice.

	Figure~\ref{fig:cameras_over_map} illustrates the relationships between frame rates and geographical locations. In this figure, a small circle indicates a high frame rate. When a high frame rate is desired, the data stream can  be sent only a short distance - measured by the round-trip time (RTT). This requires the resource manager to analyze the data stream at a cloud instance near the network camera. In Figure~\ref{fig:cameras_over_map} (a), six separate cloud instances are needed because the circles do not overlap. If a lower frame rate is acceptable, the acceptable RTT is higher and the circles can be larger, as shown in Figure~\ref{fig:cameras_over_map} (b). One cloud instance is capable of analyzing multiple data streams. As a result, only the three boxed instances are needed and the cost can be reduced. 
	% The only piece missing in the paper now is the updated figure, I will change this last line once we add the Numbers/Letters to the figure to clear up which three

\begin{figure}[h]
    \centering
    \includegraphics[width=5in]{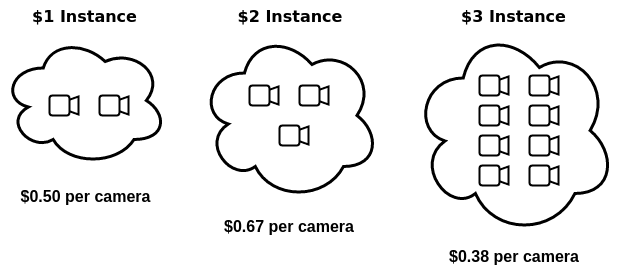}%
    \caption{There are 3 possible cloud instances, each with a different size and cost.}
    \label{fig:cameras_in_cloud}
\end{figure}
\subsubsection*{\textbf{Instance Type Optimization}}
Typically, when multiple data streams are analyzed at one instance, additional cores, memory, or the presence of GPUs are required. This results in higher costs. Thus, to effectively optimize cloud instance usage, the resource manager has to consider the number of instances as well as their capabilities. Consider  the example in Figure~\ref{fig:cameras_in_cloud}. Here, data from eight network cameras is analyzed, and the cloud manager must decide which instances to use. The cloud manager has the options to choose three types of instances at \$1, \$2, and \$3 per hour.  The first instance has the fewest cores and the least amount of memory; consequently, it can analyze only two data streams. The third type of instance, despite the higher cost, can analyze eight data streams at the lowest cost per stream.

\begin{figure}[h]
  \centering
  {\includegraphics[width=5in]{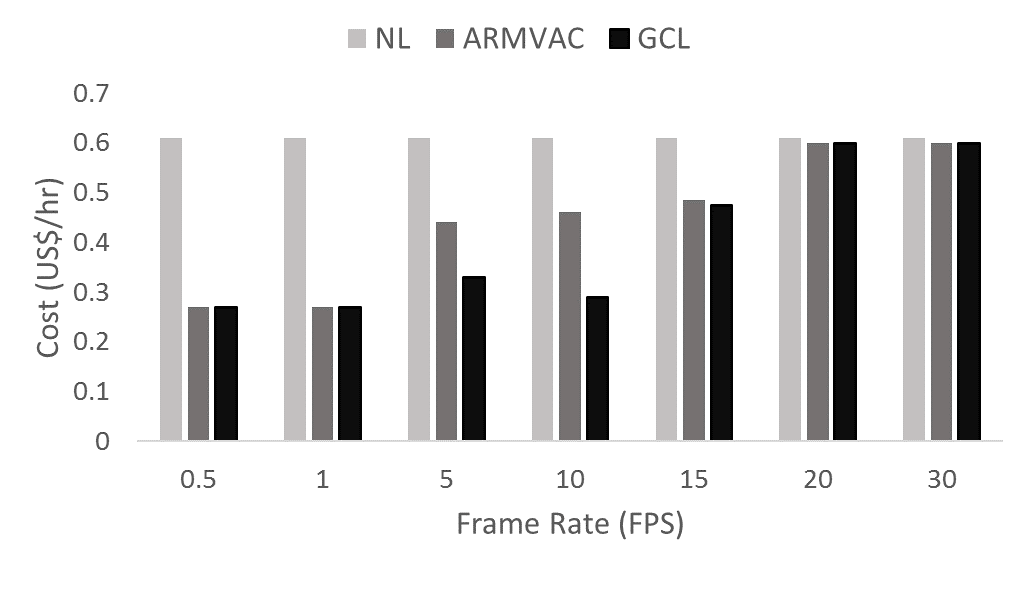}}
  \caption{Consider three different cloud resource managers: Nearest Location (NL), Adaptive Resource Management for Video Analysis in the Cloud (ARMVAC), and Globally Cheapest Location (GCL). A cost comparison between each resource manager is shown at various target frame rates.
  }
  \label{fig:costComparison_frameRates}
\end{figure}

Considering instances' types and locations simultaneously makes cloud resource management a complex optimization problem. Mohan et al.~\cite{mohan2016location, mohan2018adaptive} propose to first eliminate instance locations that are outside the acceptable RTT range. This method, named ARMVAC, then selects the lowest-cost instances from the remaining pool, and sends as many data streams to this instance while meeting the desired frame rates. This strategy performs well for high and low frame rates; streams with higher than 20 frames per second perform well since few instances can meet the processing requirements. Analyzing data streams with lower than one frame per second also performs well since there are few restrictions on instance requirements. The method does not perform well, however, when the desired frame rates are between one and twenty frames per second. In this range there are too many instance selections that can analyze the data. Mohan et al. \cite{mohan2016location} resolve this issue by formulating it as the multi-dimensional, multi-choice packing problem that accounts for the camera to cloud instance price ratio. This method, named Globally Cheapest Location (GCL), can reduce cost by as much as 56\% compared with a resource manager that always selects the Nearest Location (NL) instances, and 31\% compared with the ARMVAC method. An evaluation of the relationship between cost and frame rates is shown in Figure \ref{fig:costComparison_frameRates} which compares ARMVAC, GCL, and NL solutions. As explained earlier, the analysis programs' resource demands may vary due to a wide range of reasons.
These methods can make resource decisions quickly and be applied during runtime.  An experiment shows the adaptive solutions implemented in Amazon EC2 responding to
the changing needs is presented in~\cite{7450565}.
\label{sec:GCL_reference} 

\section*{Summary}\label{sec:summary} 

With the rise of the ``Internet of Video Things''~\cite{6410856}, comes the possibility to make use of the massive amount of visual data. Analyzing the data requires a large amount of cloud computing resources. With the variety of cloud services available, it is important to optimize cloud-instance utilization to save money. This work proposes a cloud resource manager to make cost-effective use of both the real-time video data available on the Internet and the wide variety of cloud services available. The resource manager determines cost-effective ways to analyze video streams using cloud instances. It considers the geographic location of an instance relative to a camera, as well as the  resources available in particular instances. By taking these factors into consideration, more than 50\% cost can be saved when using a commercial cloud vendor.
\section*{Acknowledgements}\label{sec:acknowledgements} 

This research project is supported by the National Science Foundation OAC-1535108, IIP-1530914,  OISE-1427808, and CNS-0958487.
We also acknowledge the Lynn CSE Fellowship at Purdue University, Amazon Web Services, Microsoft Azure, Google, Facebook, and Intel for their financial or technical supports. Thiruvathukal has a director's discretionary allocation  from the Argonne National Laboratory to support the supercomputing aspects of this research. We thank the owners of the data for the permission to conduct the experiments. Any opinions, findings, and conclusions or recommendations expressed in this material are those of the authors and do not necessarily reflect the views of the sponsors.
\section*{About The Authors}\label{sec:bios}

% \par\noindent 
% \parbox[t]{\linewidth}{
% \noindent\parpic{\includegraphics[height=1.5in,width=1in,clip,keepaspectratio]{}}
% \noindent \textbf{Zohar Kapach} is pursuing his B.S degree in computer engineering from Purdue University, West Lafayette with an expected graduation date of Fall 2019. His research interests include machine learning and computer vision.}
% \vspace{4\baselineskip}

\par\noindent 
\parbox[t]{\linewidth}{
\noindent
\begin{wrapfigure}{l}{0.25\textwidth} 
    % \vspace{-\baselineskip}
    \includegraphics[height=1.5in,width=1in,clip,keepaspectratio]{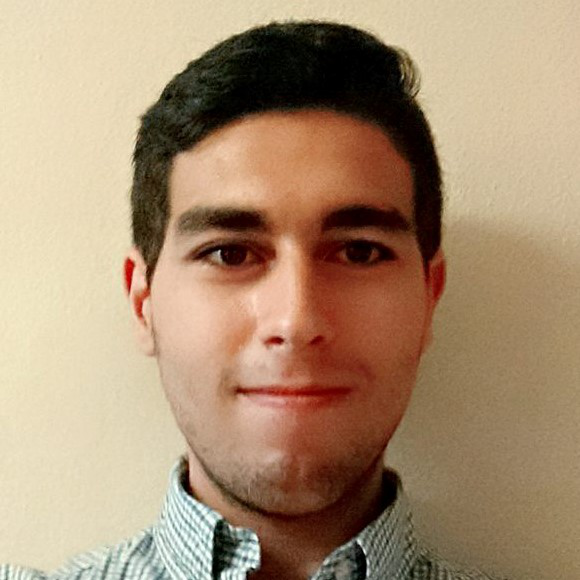}
    
\end{wrapfigure}

\noindent \textbf{Zohar Kapach} is pursuing his B.S degree in computer engineering from Purdue University. He is leading the transfer learning research team in the CAM$^2$ project. His research interests include cloud resource optimization, deep learning, and computer vision. \newline
Email: zkapach@purdue.edu }
\vspace{2\baselineskip}

\par\noindent 
\parbox[t]{\linewidth}{
\noindent
\begin{wrapfigure}{l}{0.25\textwidth} 
    \vspace{-\baselineskip}
    \includegraphics[height=1.5in,width=1in,clip,keepaspectratio]{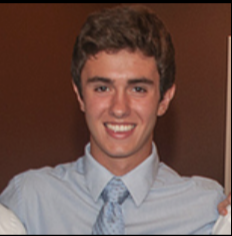}
\end{wrapfigure}
\noindent \textbf{Andrew Ulmer} is a senior undergraduate student studying computer engineering and statistics at Purdue University. His interests include deep learning, computer vision, and entrepreneurship. \newline
Email: ulmera@purdue.edu}
\vspace{2\baselineskip}

\par\noindent 
\parbox[t]{\linewidth}{
\noindent
\begin{wrapfigure}{l}{0.25\textwidth}
 \vspace{-\baselineskip}
    \includegraphics[height=1.5in,width=1in,clip,keepaspectratio]{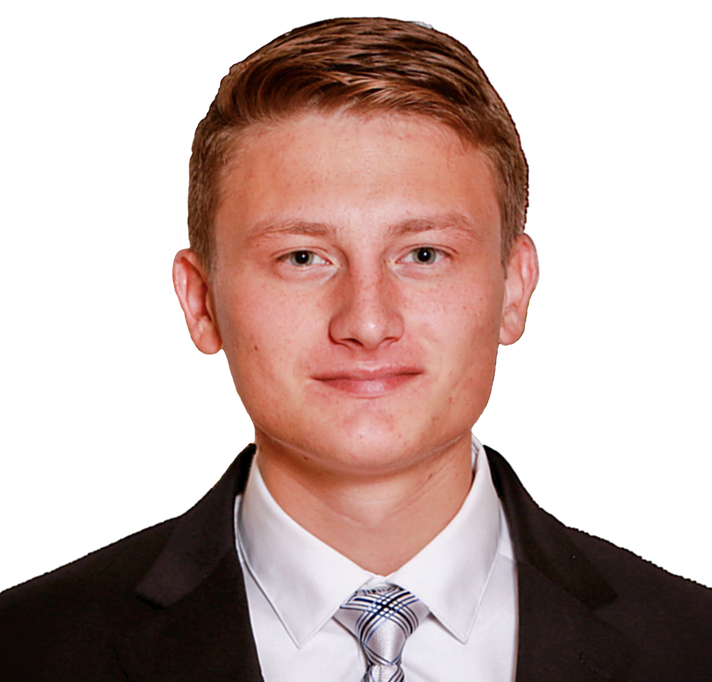}
\end{wrapfigure}
\noindent \textbf{Daniel Merrick} is pursuing his B.S degree in electrical engineering from Purdue University, West Lafayette with an expected graduation date of Fall 2019. His research interests include machine learning and computer vision. \newline
Email: dmerrick@purdue.edu}
\vspace{2\baselineskip}

\par\noindent 
\parbox[t]{\linewidth}{
\noindent
\begin{wrapfigure}{l}{0.25\textwidth} 
 \vspace{-\baselineskip}
    \includegraphics[height=1.5in,width=1in,clip,keepaspectratio]{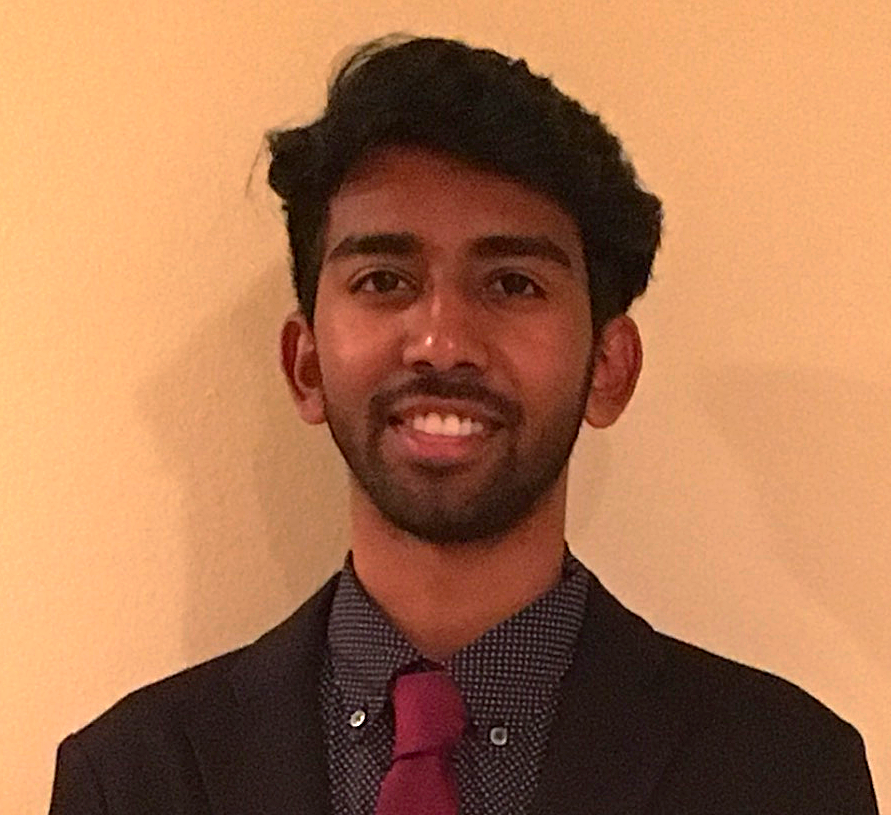}
\end{wrapfigure}
\noindent\textbf{Arshad Alikhan} is currently a BS student in Computer Science at Purdue University with a concentration in Machine Learning and a Mathematics minor. He is currently doing research with CAM2 in the area of transfer learning. His research interests include machine learning and cloud computing. \newline
Email: aalikhan@purdue.edu}
\vspace{2\baselineskip}

\par\noindent 
\parbox[t]{\linewidth}{
\noindent
\begin{wrapfigure}{l}{0.25\textwidth} 
 \vspace{-\baselineskip}
    \includegraphics[height=1.5in,width=1in,clip,keepaspectratio]{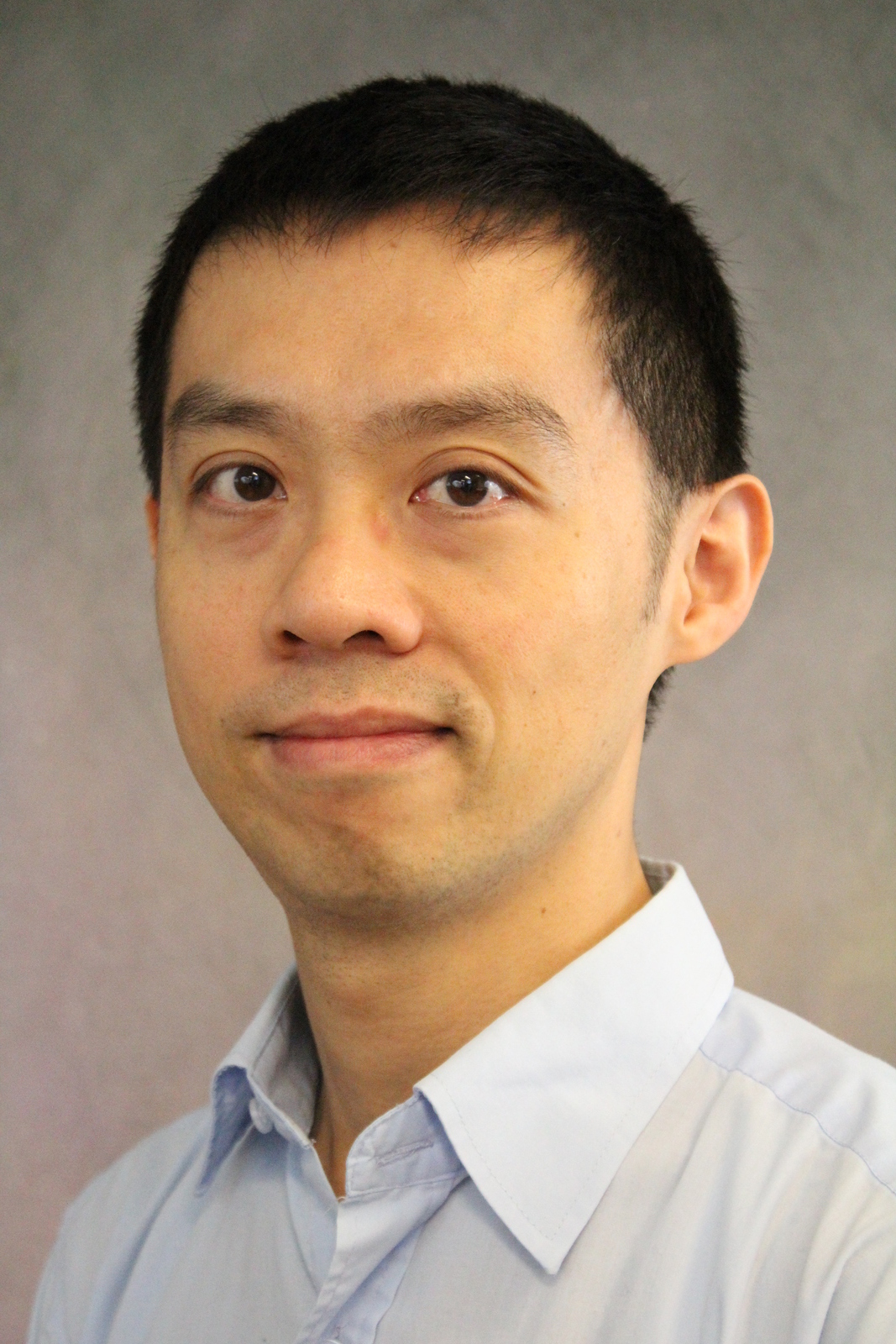}
\end{wrapfigure}
\noindent\textbf{Yung-Hsiang Lu} is a professor in the School of Electrical and Computer Engineering and (by courtesy) the Department of Computer Science of Purdue University. He is an ACM distinguished scientist and ACM distinguished speaker. Dr. Lu is a co-founder and the scientific adviser of a technology company using video analytics to improve shoppers' experience in physical stores. \newline
Email: yunglu@purdue.edu}
\vspace{2\baselineskip}

\par\noindent 
\parbox[t]{\linewidth}{
\noindent
\begin{wrapfigure}{l}{0.25\textwidth} 
 \vspace{-\baselineskip}
    \includegraphics[height=1.5in,width=1in,clip,keepaspectratio]{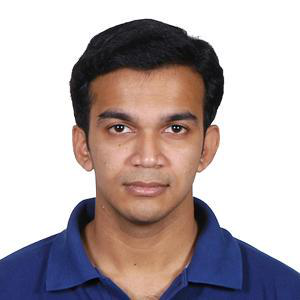}
\end{wrapfigure}
\noindent\textbf{Anup Mohan} obtained his Ph.D. from the
School of Electrical and Computer Engineering
at Purdue University in 2017. His research interests
include large-scale video analysis, cloud
computing, and big data analysis. Anup Mohan
is currently working at Intel Corporation, Santa
Clara, U.S.A. \newline
Email: anup.mohan@intel.com}
\vspace{2\baselineskip}

\par\noindent 
\parbox[t]{\linewidth}{
\noindent
\begin{wrapfigure}{l}{0.25\textwidth} 
 \vspace{-\baselineskip}
    \includegraphics[height=1.5in,width=1in,clip,keepaspectratio]{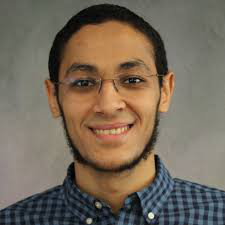}
\end{wrapfigure}
\noindent\textbf{Ahmed S. Kaseb} is an assistant professor of computer engineering in the Faculty of Engineering at Cairo University. He obtained the Ph.D. in computer engineering from Purdue University in 2016. He obtained the M.S. and B.E. in computer engineering from Cairo University in 2013 and 2010 respectively. \newline
Email: akaseb@eng.cu.edu.eg}
\vspace{2\baselineskip}

\par\noindent 
\parbox[t]{\linewidth}{
\noindent
\begin{wrapfigure}{l}{0.25\textwidth} 
 \vspace{-\baselineskip}
    \includegraphics[height=1.5in,width=1in,clip,keepaspectratio]{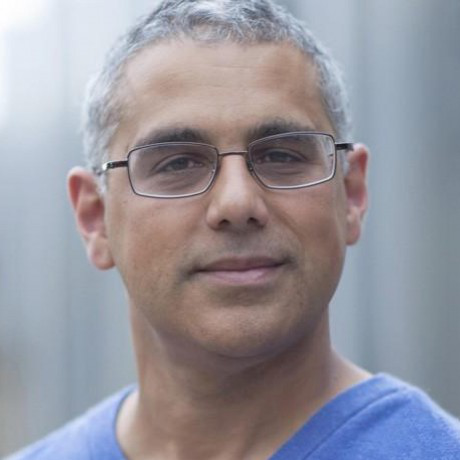}
\end{wrapfigure}
\noindent\textbf{George K. Thiruvathukal} is a Professor of Computer Science at Loyola University Chicago and visiting faculty at Argonne National Laboratory in the Argonne Leadership Computing Facility. \newline
Email: gkt@cs.luc.edu}
\vspace{2\baselineskip}

\balance
% \printbibliography
\bibliography{master}
\newpage

\end{document}